\documentclass{CRPITStyle} 
\usepackage[authoryear]{natbib}
\renewcommand{\cite}{\citep}
\usepackage{graphicx}
\usepackage{amsmath}
\usepackage{eucal}
\usepackage{enumitem}

\newtheorem{definition}{Definition}
\usepackage{caption} 
\usepackage{amsmath,amssymb}
\makeatletter
\newsavebox\myboxA
\newsavebox\myboxB
\newlength\mylenA
\newcommand*\xoverline[2][0.75]{%
    \sbox{\myboxA}{$\m@th#2$}%
    \setbox\myboxB\null
    \ht\myboxB=\ht\myboxA%
    \dp\myboxB=\dp\myboxA%
    \wd\myboxB=#1\wd\myboxA
    \sbox\myboxB{$\m@th\overline{\copy\myboxB}$}
    \setlength\mylenA{\the\wd\myboxA}
    \addtolength\mylenA{-\the\wd\myboxB}%
    \ifdim\wd\myboxB<\wd\myboxA%
       \rlap{\hskip 0.5\mylenA\usebox\myboxB}{\usebox\myboxA}%
    \else
        \hskip -0.5\mylenA\rlap{\usebox\myboxA}{\hskip 0.5\mylenA\usebox\myboxB}%
    \fi}    
\newsavebox\IBoxA \newsavebox\IBoxB \newlength\IHeight
\newcommand\TwoFig[6]{
  \sbox\IBoxA{\includegraphics[width=0.48\textwidth]{#1}}
  \sbox\IBoxB{\includegraphics[width=0.48\textwidth]{#4}}%
  \ifdim\ht\IBoxA>\ht\IBoxB
    \setlength\IHeight{\ht\IBoxB}\else\setlength\IHeight{\ht\IBoxA}\fi%
  \begin{figure*}[!htb]
  \minipage[t]{0.45\textwidth}\centering
  \includegraphics[height=\IHeight]{#1}
  \caption{#2}\label{#3}
  \endminipage\hfill
  \minipage[t]{0.45\textwidth}\centering
  \includegraphics[height=\IHeight]{#4}
  \caption{#5}\label{#6}
  \endminipage 
  \end{figure*}%
}
\graphicspath{ {images/} }
\makeatother

\begin{document}

\title{Finding Influentials in Twitter: A Temporal \\Influence Ranking Model}

\author{Xingjun Ma$^{1,2}$
\and 
Chunpin Li$^2$
\and 
James Bailey$^1$
\and 
Sudanthi Wijewickrema$^1$}
\affiliation{$^1$ Department of Computing and Information Systems \\
The University of Melbourne, \\
Victoria 3010, Australia, \\
Email:~{\tt \{xingjunm@student., baileyj@, swijewickrem@\}unimelb.edu.au}\\[.1in]
$^2$ School of Software \\
Tsinghua University, \\
Beijing 100084, China, \\
Email:~{\tt cli@tsinghua.edu.cn}}


\maketitle



\toappear{Copyright \copyright 2016, Australian Computer Society, Inc. This paper appeared at the Fourteenth Australasian Data Mining Conference, Canberra, Australia. Conferences in Research and Practice in Information Technology, Vol. 170. Yanchang Zhao, Md Zahidul Islam, Glenn Stone, Kok-Leong Ong, Dharmendra Sharma and Graham Williams (Eds.). Reproduction for academic, not-for profit purposes permitted provided this text is included.}




\begin{abstract}
With the growing popularity of online social media, identifying influential users in these social networks has become very popular. Existing works have studied user attributes, network structure and user interactions when measuring user influence. In contrast to these works, we focus on user behavioural characteristics. We investigate the temporal dynamics of user activity patterns and how these patterns affect user interactions. We assimilate such characteristics into a PageRank based temporal influence ranking model (TIR) to identify influential users. The transition probability in TIR is predicted by a logistic regression model and the random walk, biased according to users' temporal activity patterns. Experiments demonstrate that TIR has better performance and is more stable than the existing models in global influence ranking and friend recommendation.
\end{abstract}
\vspace{.1in}

\noindent {\em Keywords:} social networks, user influence, influence ranking, influentials, PageRank 

\section{Introduction}
As one of the most popular online social networks (OSNs), Twitter allows users to share status, activities and ideas, and follow others for updates. With millions of closely connected users, it has changed the way people communicate and socialize, and has had prominent influence on important historical events, such as the 2016 US presidential election and the 2010 Arabic Spring. Opinion leaders, authorities, and media hubs (or influentials) play an important role in generating and propagating such influence. With a large number of followers, they can spread stories, ideas and product information to massive audiences and enable follow-up discussions. As such, identifying influentials and understanding why and how they influence others has become a hot topic in recent years.

However, identifying influentials from millions of users is a challenging task. First of all, the definition of influence varies according to context and application domain. For example, in viral marketing, a car company may regard authoritative car buyers or advisers as influential users, while a fashion company may be more interested in pop singers. Second, it is hard to find accurate influence measures. As people's interests and behaviours change over time, the influence of a user also varies from time to time.

Existing works in this area have studied user attributes, network structure, and user interactions when measuring user influence. Different from existing works, we focus on the temporal dynamics of user activity pat-
terns and how these patterns affect user influence. In this paper, we make the following contributions: 1) we comprehensively investigate the user activity patterns and their influence on user interactions, 2) we incorporate the temporal user behaviour characteristics into a new influence ranking model to identify influential users and 3) we empirically demonstrate that our model has better performance and is more stable than the existing models in global influence ranking and friend recommendation.

The rest of the paper is organized as follows. We review related work in Section 2. Section 3 investigates user statistics, user activity patterns as well as user interaction (response) characteristics. We formulate the response probability between two users in Section 4. In Section 5, we introduce a novel influence ranking model based on the response probability. The proposed model is evaluated in Section 6. Section 7 concludes the paper and discusses future work.

\section{Literature Review}
In the areas of social science and communication theory, the study of people's individual influence in a community has been an intriguing topic since the 1940s. \citet{lazarsfeld1948peoples} introduced a two-step flow theory to formulate the part played by people in the flow of mass communications. The theory states that individuals receive media effects indirectly from opinion leaders rather than directly from mass media. In the following decades, opinion leaders or ``influentials" became an important part of innovation diffusion \citep{rogers1962diffusion}, communication theory \citep{mcquail1987mass}, and marketing \citep{chan1990characteristics}.

Modern views of user influence provide a more in-depth understanding of the diffusion process and interpersonal interactions. \citet{watts2007influentials} reported that large cascades of influence are driven not only by influentials but also by a critical mass of easily influenced individuals. In recent years, the rise of online social media has facilitated empirical exploration and validation of different influence theories.

Motivated by the design of viral marketing strategies and \citet{domingos2001mining, richardson2002mining} were the first to introduce influence modelling in online social networks from a data mining perspective. They proposed a probabilistic model to exploit the expected network value of users during the propagation of influence to others. \citet{kempe2003maximizing} investigated the propagation process of user influence in two propagation models, i.e., Linear Threshold Model and Independent Cascade Model. This work provides the first provable approximation guarantees for greedy approximation algorithms in influence maximization problem. Follow-ups in this direction are \citet{leskovec2007cost} and \citet{chen2010scalable}.

Recent works quantify user influence based on user attributes, for example, the number of followers. \citet{kwak2010twitter} reported that there is a gap in influence inferred from the number of followers and that from the popularity of one's tweets. \citet{cha2010measuring} conducted an in-depth comparison of three measures of influence: indegreee (i.e., the number of followers), retweets, and mentions. They found that popular users who have high indegree are not necessarily influential, which indicates that the number of followers may not be a good measure of influence in Twitter. This conclusion is also validated in a recent study \citep{cataldi201510}. There are other works that utilize user attributes to measure user influence such as \citet{leavitt2009new}. However, all these works demonstrate that using user attributes alone may not be accurate enough for user influence measurement.

Other works measure user influence by taking both the network structure and user interactions into consideration. One such work is TunkRank\footnote{http://thenoisychannel.com/2009/01/13/a-twitter-analog-to-pagerank/}, a variant of PageRank \citep{page1999pagerank}. The influence is calculated iteratively by the following equation:

\begin{equation*}
Influence(X) = \sum_{Y \in Follwers(X)} \frac{1+p*Influence(Y)}{|Friends(Y)|},
\end{equation*}
\noindent It measures a user's influence by the expected number of people who will read a tweet that $X$ tweets, including all retweets of that tweet. $Friends(Y)$ is the set of users that $Y$ follows while $Follwers(X)$ indicates the users that follow $X$. If $Y$ reads a tweet from $X$, there is a constant probability $p$ that $Y$ will retweet it. However, by taking $p$ a constant probability, TunkRank assumes the retweet probabilities are the same for any tweets between any users, which may not be a reliable assumption in reality.

\citet{kwak2010twitter} proposed a method to find influentials by considering both the network structure and the temporal order of information adoption. They assume that followers only adopt the information they are first exposed to and will ignore such information in later exposures. The temporal sequence of information adoption was also discussed in \citet{bakshy2011everyone} to calculate the influence score for a given URL post. Although these two works consider the chronological order of tweet diffusion, the temporal order in which information is received is not necessarily consistent with the order of information adoption. For example, users would probably read the latest relevant tweet they see when online instead of the first tweet received when offline. Therefore, the temporal pattern of user activity also affects the order of information adoption, which will be further investigated in this paper.

More in-depth investigations of user interactions have also been undertaken in existing literature. Apart from the network structure, \citet{tang2009social} also investigated the topical similarity between two users and proposed Topical Affinity Propagation (TAP) to model the topic-level social influence on large networks. A topic-based model TwitterRank was introduced by \citet{weng2010twitterrank} to estimate user influence with pre-computed topic distributions. It is also a variant of PageRank, but with topic-specific random walk. The transition probability from one user to another is defined as:

\begin{equation*} \label{eq:TR}
P_t(u,v) =\frac{\vert\tau_v\vert}{\sum_{a:\:u \: follows \: a}\vert\tau_a\vert}sim_t(u,v),
\end{equation*}

\noindent where, $\tau_v$ is the number of tweets posted by $v$, $\sum_{a:\:u \: follows \: a}\vert\tau_a\vert$ is the number of tweets posted by all fiends of $u$, and $sim_t(u,v)$ measures the topic similarity between $u$ and $v$. 

\citet{bi2014scalable} proposed a Followship-LDA (FLDA) model to integrate both topic discovery and social influence analysis in the same generative process and demonstrated it produced precise results. \citet{katsimpras2015determining} recommended a supervised algorithm, i.e., Topic-Specific Supervised Random Walks (TS-SRW), to measure user influence using PageRank with transition probability biased towards influential users. These topical models also overlooked the behavioral characteristics of the interaction itself. Dynamic user activity patterns can also impact user interactions; after all, topic interest is a rather static user profile based on long-term observations.

More recent works demonstrated new influence estimation models in continuous-time diffusion networks and multi-source social networks. \citet{du2013scalable} proposed a randomized influence estimation algorithm in continuous-time diffusion networks, which can provide a more accurate estimate of the number of follow-ups. \citet{rao2015klout} proposed a hierarchical framework to generate an overall influence score by combining user information from multiple networks and communities. In this paper, we measure user influence by incorporating user attributes, network structure, topical similarity, and user activity patterns into a PageRank based ranking model. We formulate the transition probability by the expected number of responses which is predicted by a logistic regression model, and bias the random walk towards more active users based on user temporal patterns.

\section{User Activity Analysis}
In this section, we undertake an initial analysis to reveal user statistics and temporal activity patterns so as to better understand how user activity patterns affect user influence. The social network used in this paper is a sub-network of Twitter: 7.2K New York users, 751K following links and 3.5M tweets (including 565K retweets and 787K replies). 

The sub-network was collected through Twitter Streaming API\footnote{https://dev.twitter.com/streaming/overview} with region of interest specified to New York City. The API returns approximately 1\% of randomly sampled real-time data (tweets and users) with respect to our region-specific query \cite{morstatter2013sample}. In this step, 10K New York users were collected. We then extracted the largest connected component (7.5K users) of this sampled network. Finally, we collected these users' profiles and tweets between 24th December 2013 and 24th January 2014 via the REST API \footnote{https://dev.twitter.com/rest/public}. As users that were extremely inactive have little contribution to the influence analysis, those users with less than 20 tweets were removed to get our final dataset.

\subsection{User Statistics}
To better understand user statistics, we illustrate the distribution of the number of users over the number of followers, the number of friends, and the number of tweets in Figures \ref{fig:numOfFollowers}, \ref{fig:numOfFriends}, and \ref{fig:numOfTweets} respectively. As can be seen from the figures, the three distributions all follow a power law distribution which is consistent with previous observations \citep{ritterman2009using, kwak2010twitter, petrovic2011rt}. Figure \ref{fig:followerOfFriends} shows a positive correlation between the number of followers and the number of friends, which indicates that a user with more friends tends to have more followers.

\TwoFig{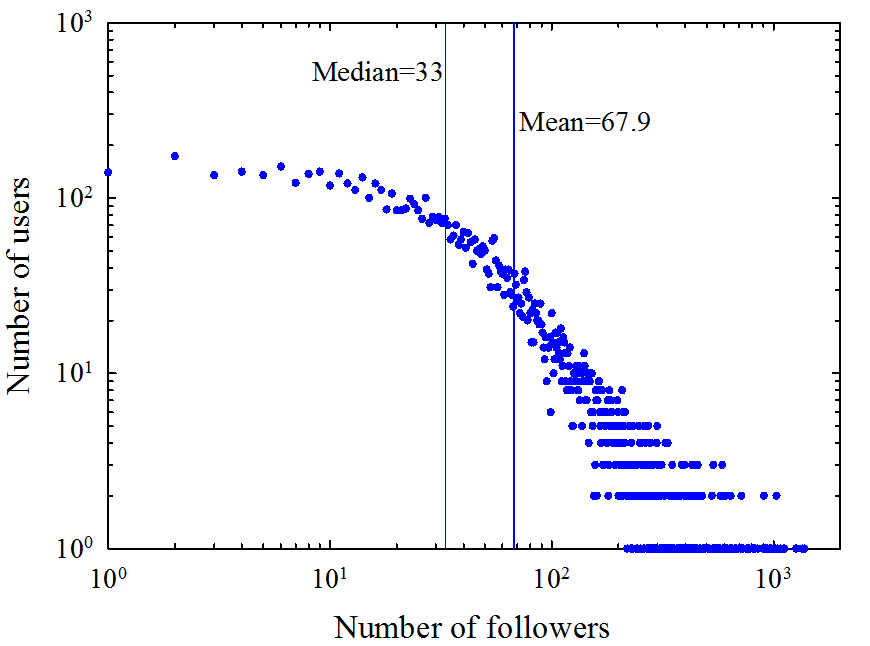}    {Number of users vs. Number of followers}    {fig:numOfFollowers}
       {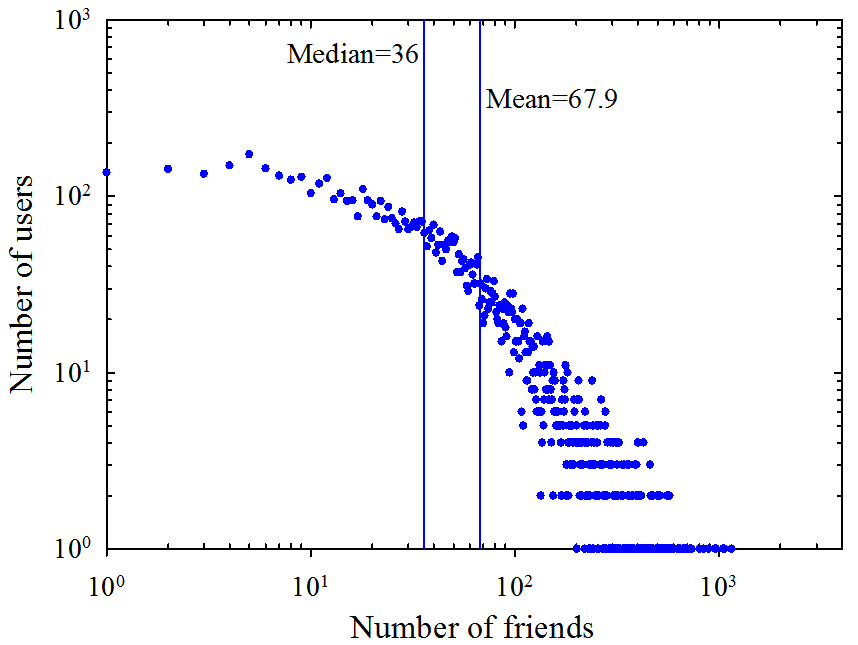}{Number of users vs. Number of friends}    {fig:numOfFriends}

\TwoFig{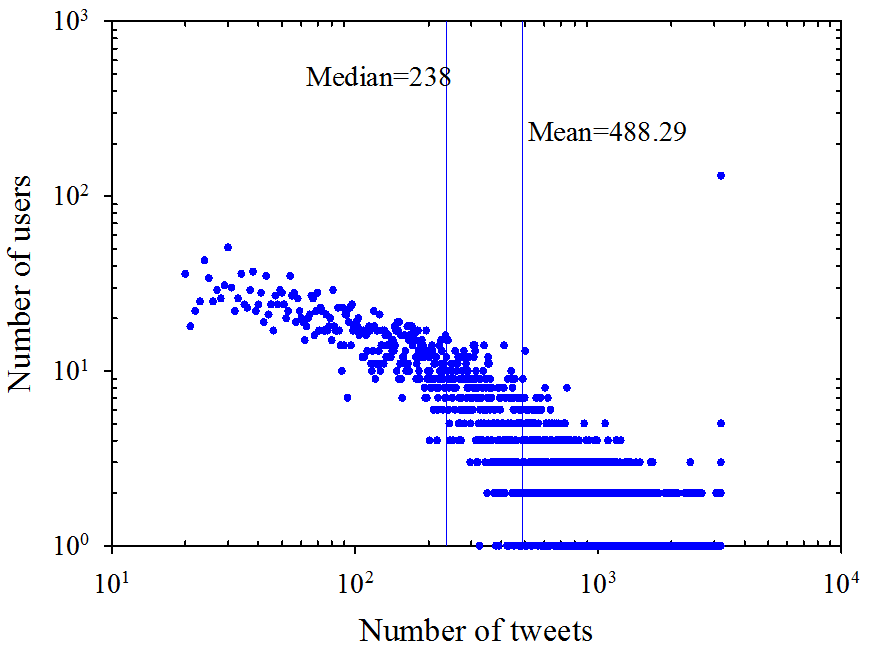}    {Number of users vs. Number of tweets}    {fig:numOfTweets}
       {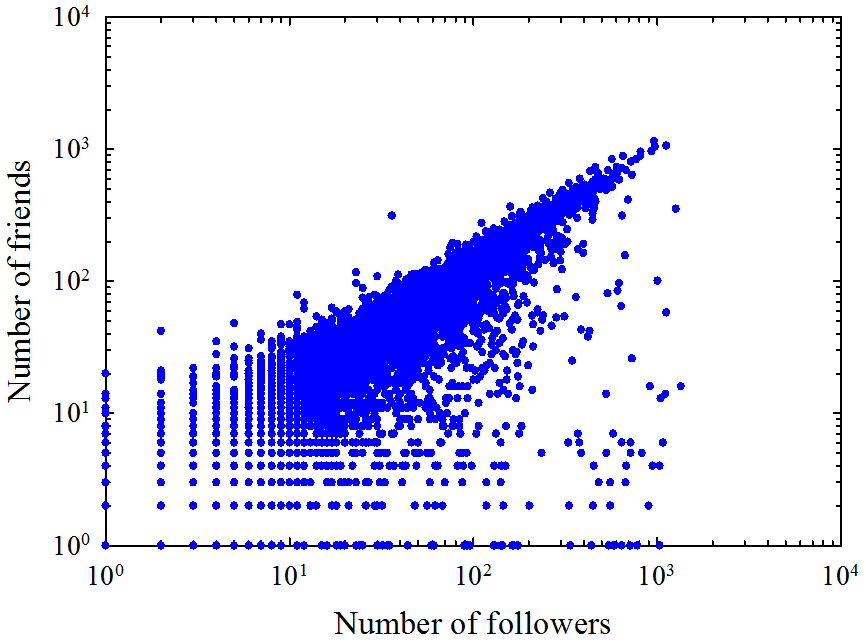}{Number of followers vs. Number of friends}    {fig:followerOfFriends}

\subsection{Temporal User Activity Patterns}
As tweet, retweet, and reply are the three major user activities, we approximate the overall user activity in Twitter by the total number of these three activities of all users. The overall user activity pattern is revealed at two different time granularities: hour of day and day of week. As shown in Figure \ref{fig:activityOfHour}, users are more active from 10:00 a.m. to 23:00 p.m. than in other hours which is consistent with people's daily work-rest routine. The weekly pattern illustrated in Figure \ref{fig:activityOfDayOfWeek} reveals that users are more active on Monday and Tuesday and less active on Friday and Saturday. This may imply that people find it hard to focus on their work on Monday and Tuesday after the weekend. A more detailed user activity pattern is depicted in the form of a heat map in Figure \ref{fig:activityHeatMap}.

\TwoFig{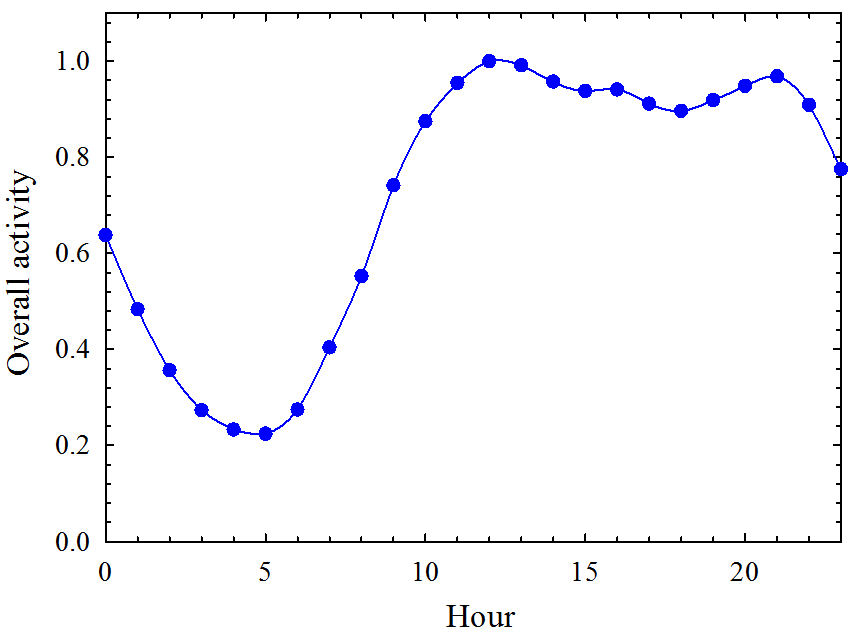}    {Hourly user activity pattern}    {fig:activityOfHour}
       {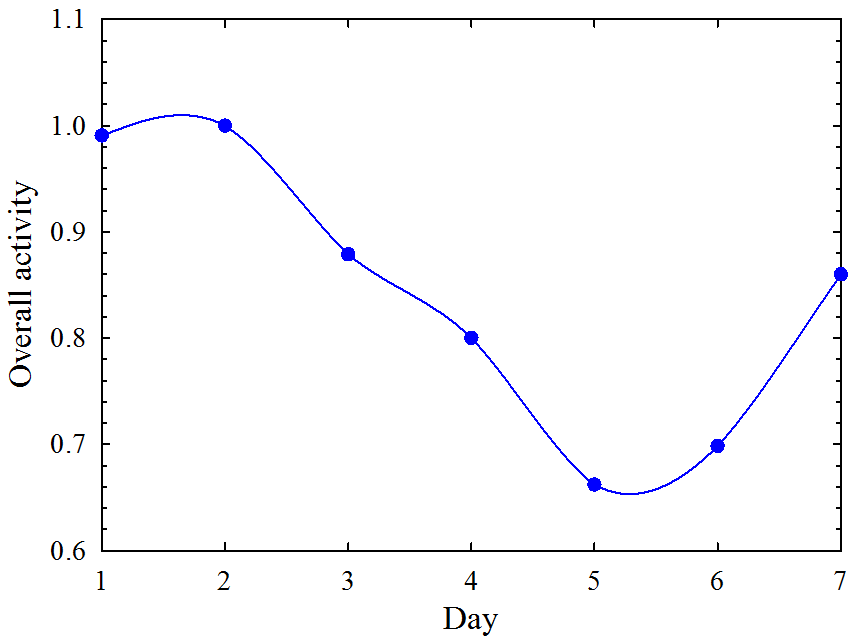}{Weekly user activity pattern}    {fig:activityOfDayOfWeek}

To answer the question of whether everybody has similar activity patterns, we further investigate individual-level activity patterns by clustering them into several clusters using the K-Spectral Centroid (K-SC) algorithm \citep{yang2011patterns}. K-SC is a shape-based clustering algorithm derived from K-Means and invariant to scaling and shifting. The optimal number of clusters can be found by the Average Silhouette Coefficient (ASC) which measures the intra-cluster cohesion and inter-cluster separation \citep{rousseeuw1987silhouettes}. Figure \ref{fig:3Clusters} illustrates the three most common patterns ($k=3$) along with the proportion of users in each cluster.

As can be seen from Figure \ref{fig:3Clusters}, the three patterns (especially $C_3$) are very different to each other. More specifically, $55\%$ of users ($C_1$ and $C_2$) are more active between 14:00 p.m. and 21:00 p.m. while the remaining $45\%$ ($C_3$) are more active between 0:00 a.m. to 4:00 a.m. Moreover, $13\%$ ($C_2$) of users are active for a shorter period of time when compared to other users and they have a significant activity burst at around 17:00 p.m. Overall, it suggests that users follow certain activity patterns and posting tweets when followers are very active may have a better chance of attracting attention.

\TwoFig{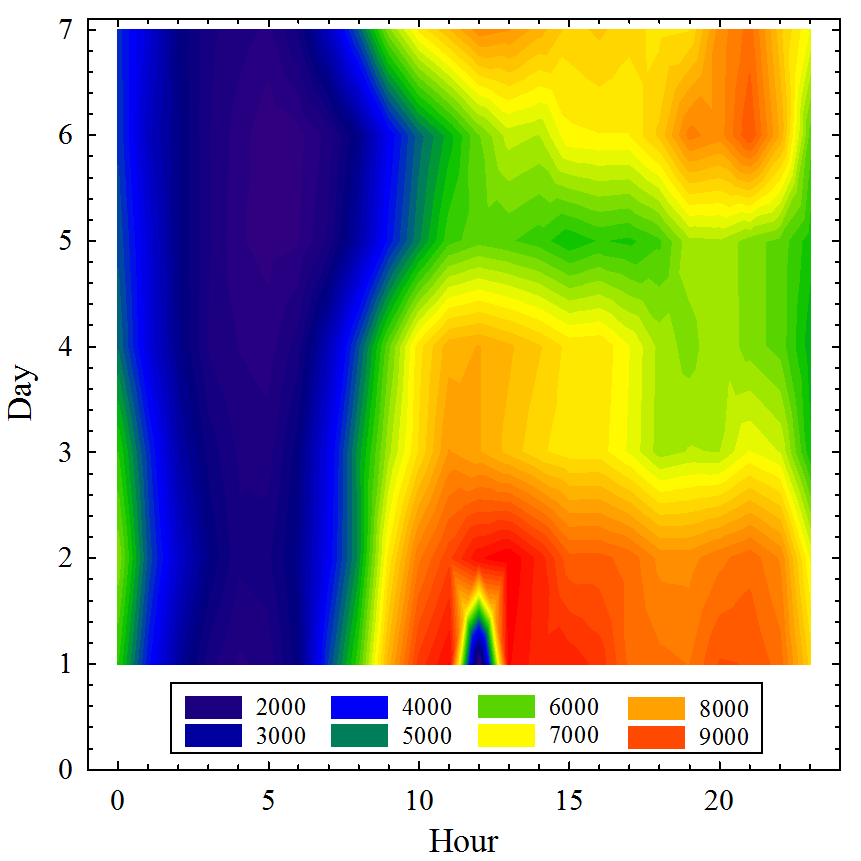}    {Heat map of user activity}    {fig:activityHeatMap}
       {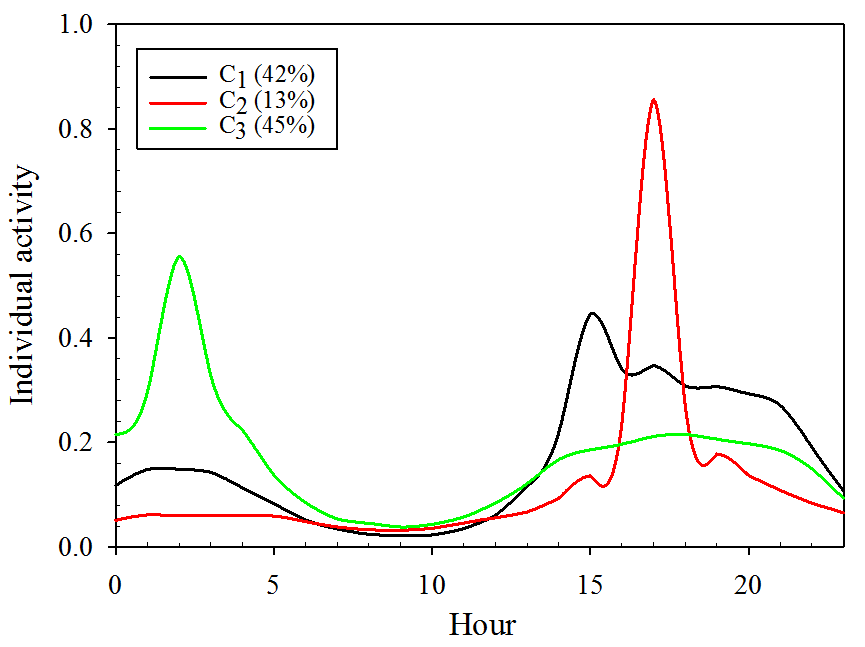}{Three common activity patterns}    {fig:3Clusters}

\subsection{Response Behaviour Analysis}
In order to provide a more in-depth understanding of how user activity patterns affect user influence, we further explore user behavioral patterns in response activity (including retweet and reply) by considering two metrics: $delay$ and $trace$. Suppose user $v$ posted a tweet $tw$ at time $t_i$, and at time $t_j$, follower $u$ responded to this tweet. Then, the two metrics can be defined by Definition \ref{def:delay} and \ref{def:trace}.

\begin{definition}
\label{def:delay}
The $delay$ in a response is the time interval between the tweet and the response.
\begin{displaymath}delay=t_j-t_i\end{displaymath}
\end{definition}

\begin{definition}
\label{def:trace}
The $trace$ is the number of earlier tweets that the follower $u$ needed to trace back before reading the tweet $tw$.
\begin{displaymath}trace=\vert\lbrace tw_k | tw_k \in RC_u \quad and \quad t_i<t_k<t_j \rbrace\vert ,
\end{displaymath}

\noindent where, $RC_u$ represents all the tweets received by $u$.
\end{definition}

The cumulative distributions of $delay$ and $trace$ of all responses are illustrated in Figures \ref{fig:delay} and \ref{fig:trace} respectively. As shown, $72\%$ of the retweets and $83\%$ of the replies occurred within 1 hour after the original tweets were posted, while $78\%$ of the retweets and $85\%$ of the replies had been done by tracing back less than $100$ earlier tweets. This indicates that users dislike reading many earlier tweets. Therefore, response activity is time-sensitive and this significantly affects users' influence on each other.

\TwoFig{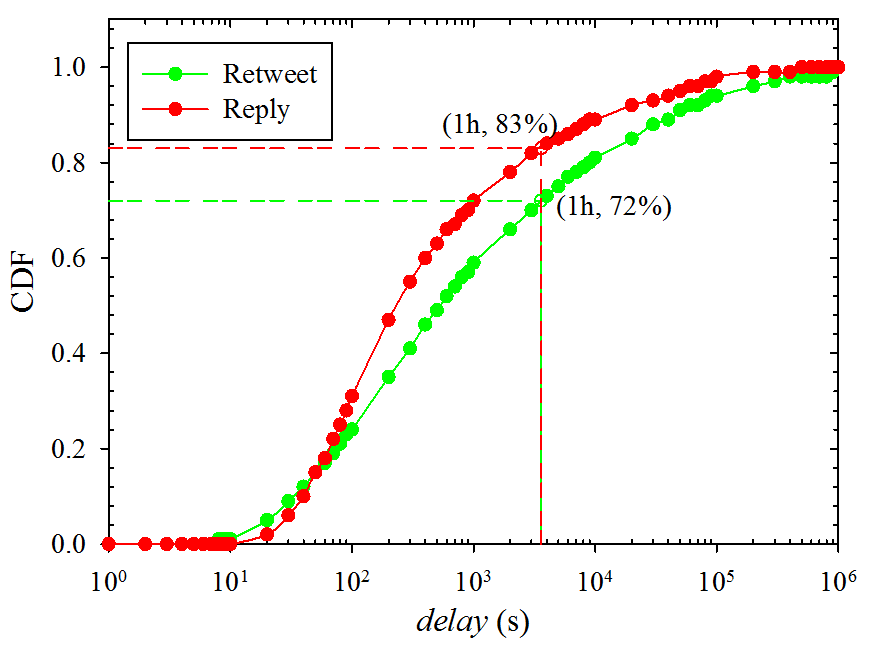}    {The cumulative distribution of delay}    {fig:delay}
       {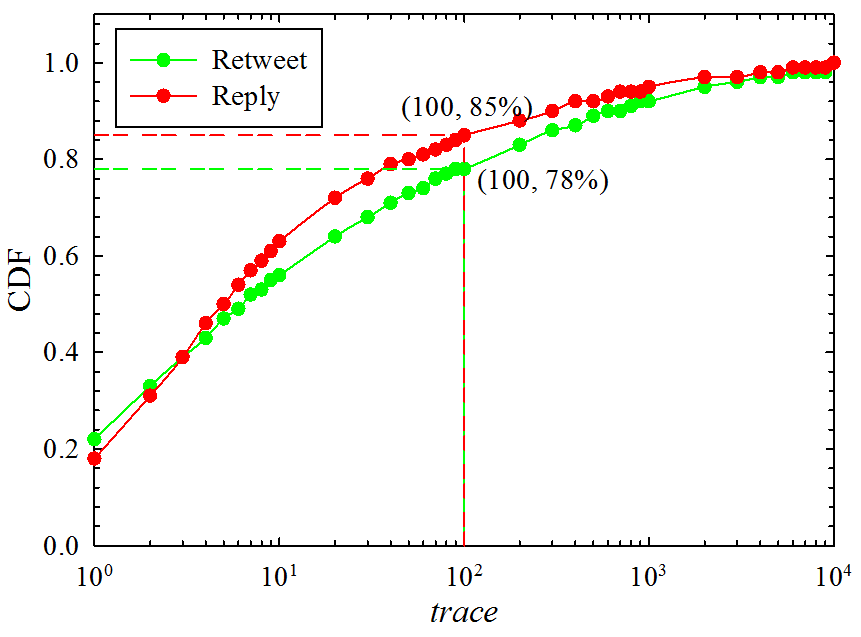}{The cumulative distribution of trace}    {fig:trace}

\section{Response Prediction and Formulation}
As response is an explicit indication that a follower has read the tweet and shows a strong interest to interact, we use responses to measure the inter-user influence. We assume that if a user responds to any tweets posted by a friend, that the user is influenced by this friend. In order to formulate the correlation between response and many other factors such as the number of friends, the number of followers, and activity patterns, we apply predictive models to predict the probability of response for a specific tweet. Frequently used notations are given in Table \ref{table:notations}.

\begin{table}[ht]
\centering
\renewcommand{\arraystretch}{1.2}
\caption{Symbolic notations}
\label{table:notations}
\begin{tabular}{c|l} \hline 
 Notation & Description \\ \hline \hline
 $V$ & user set ($u,v \in V$) \\
 $E$ & following links ($u$ follows $v$) \\
 $F$ & friend set \\
 $\mathcal{F}$ & responded friends \\
 $FL$ & follower set \\
 $T$ & posted tweets \\
 $RC$ & received tweets \\
 $RT$ & retweets  \\
 $m$ & the number of favourites \\ \hline
 \end{tabular}
\end{table}

\subsection{Feature Selection}
Features used for response prediction are selected based on our findings in user activity analysis, as well as existing works \citep{petrovic2011rt, guille2012predictive, cossu2016review}. Our 10 selected features can be grouped into 3 categories: user attributes, temporal activity, and topical similarity. 

\begin{description}
  \item[User Attributes:] \hfill
  \begin{enumerate}[leftmargin=-.1in]
  \item The number of listed times ($LI$) \hfill \\
  		According to \citep{petrovic2011rt}, this factor is more powerful in measuring popularity than the number of followers or friends.
  \item Favourites received per tweet ($FV$)  \hfill \\
  \begin{equation}
  FV_v = \frac{m_v}{|T_v|}
  \end{equation}
  \item Verified user ($VR$) \hfill \\
  \begin{equation}
  VR_v = \begin{cases} 1 &\mbox{if } \ v \ is \ verified \\ 
0 & \mbox{otherwise }  \end{cases}
  \end{equation} \\
  		It has been reported that users are more likely to reponse to verified friends \citep{petrovic2011rt}. Verified user has a blue badge next to the name indicating that this account of public interest is authentic \footnote{https://support.twitter.com}.
  		\item Retweet ratio ($RR$) \hfill \\
  \begin{equation}
   RR_v = \frac{|\mathcal{T}_v|}{|T_v|}
  \end{equation} \\
  		This feature reflects how actively a user gets involved in the interaction. The higher the $RR$, the more active the user \citep{guille2012predictive}.
  		\item Ever responded ($RE_{uv}$) \hfill \\
  \begin{equation}
  RE_{uv} = \begin{cases} 1 &\mbox{if } v \in \mathcal{F}_u \\ 
0 & \mbox{otherwise }  \end{cases}
  \end{equation} \\
  		Previous interaction records can be an indicator of how closely two users are related. Here, we define $v$ as a close friend of $u$ if $RE_{uv}=1$. Otherwise, $v$ is a normal friend of $u$.
  		\item Proportion of tweets ($PT_{uv}$) \hfill \\
  \begin{equation}
 PT_{uv} = \frac{|T_v|}{\sum_{f \in F_u}{|T_f|}}
  \end{equation} \\
  		$PT_{uv}$ is the proportion of $v$'s tweets to all tweets of $u$'s friends. The higher the $PT_{uv}$ is, the more easily $v$ can draw attention from $u$ \citep{cossu2016review}.
\end{enumerate}
  
  \item[Temporal Activity:] \hfill
   \begin{enumerate}[leftmargin=-.1in]
  \item The number of tweets posted in hour $t$ ($N^{t}$) \hfill \\
  \begin{equation}\label{eq:n_o_t}
  N_{v}^{t} = \frac{|T_{v}^{t}|}{d_v}, \ t \in [0,23]
  \end{equation}
  
  \noindent where, $d_v$ is the number of available days in our observations and is calculated by the time interval between the first and last tweet of $v$ that available in our dataset.
  \item User activity at time $t$ ($A^{t}$) \hfill \\
  \begin{equation}
  A_{v}^{t} = \frac{|N_{v}^{t}|}{\sum_{h=0}^{23}{N_{v}^{h}}}, \ t \in [0,23]
  \end{equation}
   $A^t$ represents the probability a user is active in hour $t$.
  	\item Joint activity at time $t$ ($JA^{t}$) \hfill \\
  \begin{equation}
   JA_{uv}^{t} = A_{u}^{t}A_{v}^{t}, \ t \in [0,23]
  \end{equation}
  	$JA_{uv}^{t}$ is the probability that two users are both active in hour $t$ under the assumption that $u$ and $v$ are independent.
\end{enumerate}
  \item[Topical Similarity:] \hfill
  \begin{enumerate}[leftmargin=-.1in]
  \item Topic similarity ($TS$) \hfill \\
 	\begin{equation} \label{eq:ts}
	TS_{uv} =\sqrt{2*D_{JS}(u,v)}
	\end{equation}
	\noindent $D_{JS}(u,v)$ is the Jensen–Shannon divergence between the topic distributions of two users \citep{weng2010twitterrank}.
\end{enumerate}
\end{description}


\subsection{Response Prediction}
Given two users $u$ and $v$ ($u$ follows $v$), with $v$ having posted a tweet at time $t$, the task of response prediction is to predict the probability that $u$ will respond to this tweet. For each tweet, we pair the author with each of his followers to generate an instance. This results in 2.9M instances of which $1.3\%$ are responses and the rest are non-responses. We further subsample a balanced dataset with 66K instances (including 33K responses and 32.9K non-responses) and normalize all features to $[0,1]$. It is worth mentioning that those tweets with author or follower not in the user set are removed initially.

Based on this dataset, we apply logistic regression (LR), C4.5 and multilayer perceptron (MLP) with one hidden layer (100 nodes) with 5-fold cross validation to predict the probability of response. The accuracy performance is reported in Table \ref{table:response_prediction}. As can be seen, C4.5 achieves the best accuracy followed by LR and MLP with similar performance results. 

\begin{table}[ht]
\centering
\renewcommand{\arraystretch}{1.2}
\caption{The accuracy of response prediction}
\label{table:response_prediction}
\begin{tabular}{c|c|c|c} \hline
 Classifier & LR & C4.5 & MLP \\ \hline
 Accuracy & $86\%$ & $89\%$ & $86\%$ \\ \hline
 \end{tabular}
\end{table}

\begin{figure}[!htb]
\includegraphics[width=0.45\textwidth]{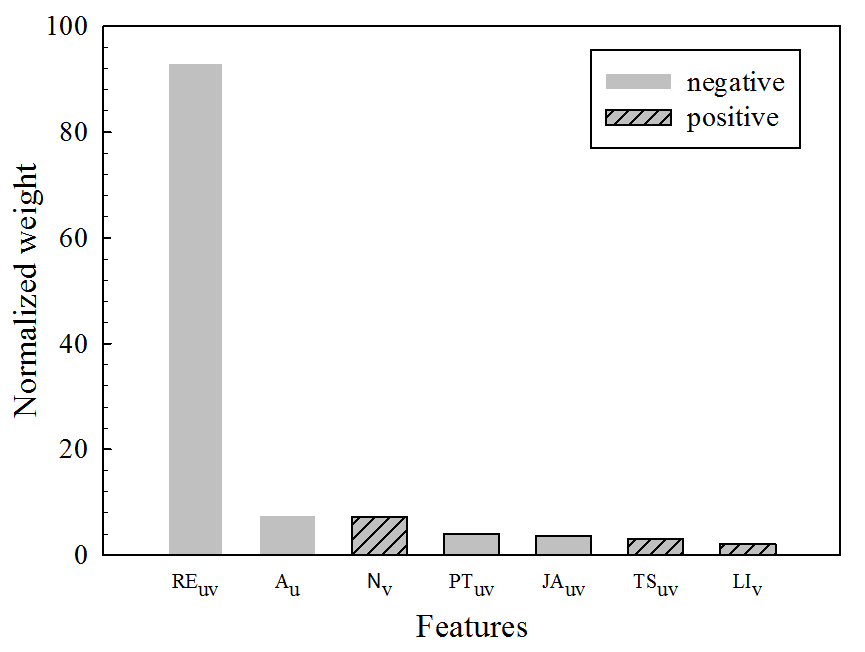}
\caption{Important features ranked by weight}
\label{fig:features_rank}
\end{figure}

Figure \ref{fig:features_rank} illustrates a few important features ranked by the absolute value of the weight learnt by LR. Feature $RE_{uv}$ (ever responded) dominates the model, in fact, nearly $60\%$ of the responses occurred between users who interacted with each other more than once. It proves that users are more prone to respond to close friends than normal friends.

For the author, the number of tweets ($N_v$) posted is the most influential factor on response. Since the weight of $N_v$ is positive, posting more tweets will, however, decrease response probability. This is because the predicted probability is averaged to one tweet which decreases when the number of tweets increases. For the follower, the normalized activity ($A_u$) with a negative weight is the most important factor. It means that the more active the follower is, the more chance there is that the tweet will get a response. This can also be prove by joint activity ($JA_{uv}$) with a positive weight. Moreover, we find that user attributes such as the number of favourites ($FV_v$), verified user ($VR_v$), and retweet ratio of the follower ($RR_u$) have little impact on response interaction.

\subsection{Response Probability Formulation}
Although LR is not the best model, the response probability of LR can be easily described in an explicit form as such can be easily integrated into an influence ranking model. There are two benefits of such integration. First, the response probability can be estimated within the influence model, otherwise, it has to be estimated separately by a prediction model. Second, it makes the influence model more flexible in adjusting the weights of different features in different application contexts.

Therefore, we adopt LR to formally define the response probability. Given feature space $S$, follower $u$, friend $v$ and a tweet posted by $v$ at time $t$, the probability that $u$ will respond to this tweet can be defined by:
\begin{equation} \label{eq:p_t}
P_{uv}^{t} =\frac{1}{1+exp(w_{0}+\sum_{i=1}^{|S|}w_iS_{i}^{t})}
\end{equation}

\section{A Temporal Influence Ranking Model}
In this section, we introduce a new temporal influence ranking (TIR) model based on PageRank model to estimate user influence so as to identify influentials. TIR is a PageRank based graphic algorithm taking the temporal homogeneity of user activity patterns into consideration. As a response is an explicit indication of user influence of one user on another, we approximate the influence of a friend to a follower by the expected number of responses the friend may receive from the follower. Since response activity is time-sensitive, this approximation is performed for each hour based on integrated hourly response probability between users.

\subsection{Model Formulation}
For the purpose of simplicity, we denote the directed graph formed by users and following links as $G(V,E)$, where $V$ is the node set with each node being a user, $E$ is the edge set with each edge being a directed link pointing from follower to friend. User influence propagates from one node to another with a certain probability along the edges in the graph. 

Given the response probability of a particular tweet, the hourly response probability can be calculated by aggregating the probability of all tweets posted during hour $t$. To better differentiate normal friend and close friend, we first fix the value of feature $RE_{uv}$ in Equation \eqref{eq:p_t} to 1, then introduce a penalty factor $c$ to penalize normal friends. Thus, the transition probability from $u$ to $v$ in hour $t$ can be defined as:

\begin{equation} \label{eq:new_p_t_h}
\mathcal P_{uv}^{t} =
\begin{cases}
cN_{v}^{t}P_{uv}^{t} &\mbox{if } \ v \in \mathcal{F}_u \\ 
(1-c)N_{v}^{t}P_{uv}^{t} & \mbox{otherwise }
\end{cases}
\end{equation}

\noindent where, $P_{uv}^{t}$ is the response probability of one tweet as defined in Equation \eqref{eq:p_t} with learnt weights and $N_{v}^{t}$ is the number of tweets posted by $v$ in hour $t$ as defined in Equation \eqref{eq:n_o_t}. $c\in[0.5,1]$ is the penalty factor that defines how close friends and normals friends are treated. $c=0.5$ indicates that close friends and normal friends are considered to be the same while $c=1$ means normal friends will be considered strangers.

Then we plug the hour-based response probability $\mathcal P_{uv}^{t}$ into the PageRank model where user influence can be calculated iteratively by the following Equation: 

\begin{equation} \label{eq:tir}
\vec{R^t}=M \times \vec{R^t},
\end{equation}

\noindent where, $M$ is a $|V|\times|V|$ transition probability matrix with element $M_{uv}$ defined as the response probability of user $u$ to user $v$ in hour $t$:

\begin{equation} \label{eq:transprob}
M_{uv}=
\begin{cases}
\mathcal P_{uv}^{t} &\mbox{if } \ u \ follows \ v \\
0 &\mbox{otherwise }
\end{cases}
\end{equation}

Equation \eqref{eq:tir} is equivalent to an eigensystem with an eigenvalue of 1. Therefore, it has a solution if and only if it meets three conditions: 1) M is a stochastic matrix, 2) M is irreducible (i.e., M is a strongly connected matrix) and 3) M is non-cyclical \cite{page1999pagerank}. However, these conditions are not satisfied in reality: 1) condition 1 is not satisfied as users with no followers will end up with zero column vectors in M, 2) condition 2 might not be satisfied as the strong connectivity of M is not guaranteed, and 3) cyclical nodes are highly possible in social networks.

To meet the three conditions, transformations should be performed on transition matrix $M$. First, we normalize $M$ by the sum of each column to make it scale invariant. Then, we add a $|V| \times |V|$ matrix with all elements of the same value $\frac{1}{|V|}$ to $M$ following Equation \eqref{eq:transformation}. The damping factor $\gamma$ represents the probability that a stranger randomly jumps to a user without following the links. As generally assumed, $\gamma$ in this paper is set to 0.85.

\begin{equation} \label{eq:transformation}
\mathcal{M}=\gamma M + (1-\gamma) \times \bigg[\frac{1}{|V|}\bigg]_{|V| \times |V|}
\end{equation}

All three of the above conditions are satisfied after these transformations. As each entry in $\mathcal{M}$ is a non-negative real number and the sum of each column equals to one, $\mathcal{M}$ is a column-stochastic matrix and also a strongly connected matrix. Since each node in $\mathcal{M}$ is connected to other nodes directly, the shortest path of each node to itself equals to one which means $\mathcal{M}$ is also a non-cyclical matrix. Thus, the proposed TIR model can be written in a new form as defined in Equation \eqref{eq:alternative}.
\begin{equation} \label{eq:alternative}
\vec{R^t}= \gamma M \times \vec{R^t} + (1 - \gamma) \times \bigg[\frac{1}{|V|}\bigg]_{|V| \times 1}
\end{equation}

Overall, by utilizing the expected number of response between two users as the transition probability, we bias the random walk in TIR towards more interactive friends and such bias is based on user attributes, temporal activity and topical similarity. We also introduce a penalty factor to further control such bias. 

\subsection{Discussion}
The TIR model has three advantages over existing models. First, it provides more accurate influence estimation. This is because people's interests and behaviours change over time. Such changes can be easily captured by TIR as it evaluates user's influence based on the dynamic information that occurred in a short period of time (days or weeks). As it doesn't rely on the whole Twitter dataset, this also makes it more efficient. Existing models such as TwitterRank depends on massive tweets to get a good estimation. Third, TIR is more flexible. This is because it incorporates a explicit formulation of logistic regression as part of the model to formulate the transition probability. Such formulation can be easily adjusted in different application contexts. For example, if the number of followers is considered to be very important in a context, we can adjust its weight in the formulation accordingly.

In Twitter, the following relationship between follower and friend is asymmetrical (or weak), that is, the friend doesn't have to follow back to his followers. However, in some other social networks such as Facebook, the following relationship is symmetrical (or strong), i.e., the friend has to accept the connection request and follow back. Our model can be easily generalized to those symmetrical networks by taking the symmetrical network as a special case of the asymmetrical network where all friends and followers follow each other.

\section{Experiments and Results}
To better evaluate TIR, we conduct two different types of experiments to contrast it with existing TunkRank and TwitterRank models. First, we compare their similarities and differences in global influence ranking. Then, we evaluate their performance in a friend recommendation task. It is worth mentioning that the experiments are based on our sampled dataset, not the whole Twitter dataset.

\subsection{Global Influence Ranking}
The user influence obtained from TIR is an hour-based influence. The global influence can be further calculated via Equation \eqref{eq:aggregation} where $w_t$ is the weight of user influence in hour $t$. $w_t$ can be the overall user activity, in which case the aggregated influence serves as the global influence. Observe that $w_t$ can also be individual activity of a particular user, and in that case, the outcome can be interpreted as the personal perspective influence.

 \begin{equation} \label{eq:aggregation}
\vec{R}=\sum_{t=0}^{23} w_t \vec{R_t}
\end{equation}

We compare the TIR model with the existing TunkRank and TwitterRank models to examine their similarities and differences in global influence ranking. The penalty factor $c$ of TIR model is set to 0.5, 0.85 and 1 in order to explore its influence to the final ranking. The top 10 influentials are listed in Table \ref{table:global_ranking}. As we can see, the top 10 influentials are of different types such as news media (e.g., ``ABC", ``nprnews and ``latimes"), food (e.g., ``WholeFoods" and ``deverfoodguy"), sports (e.g., ``HPbasketball", ``BlkSportsOnline" and ``YogaArmy") and public figures (e.g., ``chrisbrogan" and ``bomani\_jones"). It's not surprising that news media outnumbers the other types of users as latest news stories are usually well-written and very attractive.

Comparison between $TIR_{c=0.5}$ and $TIR_{c=0.85}$ indicates that TIR with a larger penalty factor makes users with more responsive followers stand out which is an advantage over other models. Take ``bomani\_jones" for example, $12\%$ of his tweets were retweeted by others which even includes top influentials such as ``talkhoops" and ``rodimusprime". ``MySOdotCom", however, only has $5\%$ retweeted tweets. When the penalty increases, the influence of ``bomani\_jones" increases while that of ``MySOdotCom" decreases.

Furthermore, we calculate the commonly used Kendall’s $\tau$ rank correlation coefficient \citep{knight1966computer} to measure rank correlations between different models. As shown in Table \ref{table:k_tau}, $TIR_{c=0.5}$ is more similar to the other two models when compared with $TIR_{c=1}$. This is because $TIR_{c=0.5}$ ignores the difference between close friend and normal friend as both TunkRank and TwitterRank do. Also, TIR is more similar to TwitterRank than to TunkRank whether $c=0.5$ or $c=1$ which is because both TIR and TwitterRank take topical similarities into consideration.

Overall, with a controllable penalty factor, TIR is more flexible than TunkRank and TwitterRank in global influence ranking. Particularly, TIR with a larger penalty factor can better differentiate those influentials with more responsive followers.

\begin{table*}[ht]
\centering
\renewcommand{\arraystretch}{1.2}
\caption{Top 10 globally ranked users}
\label{table:global_ranking}
\begin{tabular}{c|l|l|l|l|l} \hline
 Rank & $TIR_{c=0.5}$ & $TIR_{c=0.85}$ & $TIR_{c=1}$ & TwitterRank & TunkRank \\ \hline
 1 & WholeFoods & WholeFoods & bomani\_jones & XboxSupport & PATisDOPE \\
 2 & MySOdotCom & bomani\_jones & InTheBleachers & Foxmental\_X & nprnews \\
 3 & nprnews & MySOdotCom & ABC & MySOdotCom &  bigmarkspain \\
 4 & bomani\_jones & greensboro\_nc & wsbtv & denverfoodguy & MySOdotCom \\
 5 & greensboro\_nc & nprnews & NBCNews & chrisbrogan & chrisbrogan \\
 6 & XboxSupport & ABC & rodimusprime & CHRISVOSS & FastCompany \\
 7 & denversolarguy & InTheBleachers & ajc & PATisDOPE & CoryBooker \\
 8 & denverfoodguy & XboxSupport & HPbasketball & denversolarguy & WholeFoods \\
 9 & jilevin & SpitToonsSaloon & talkhoops & nprnews & latimes \\
 10 & YogaArmy & PRideas & BlkSportsOnline & bigmarkspain & ABC \\ \hline
 \end{tabular}
\end{table*}

\begin{table}[ht]
\centering
\renewcommand{\arraystretch}{1.2}
\caption{Kendall $\tau$ rank coefficient}
\label{table:k_tau}
\begin{tabular}{l|c} \hline
 Pairs & $\tau$ \\ \hline
 TwitterRank vs TunkRank & 0.49 \\ \hline
 TIR$_{c=1}$ vs TunkRank & 0.28 \\
 TIR$_{c=0.5}$ vs TunkRank & 0.51 \\ \hline
 TIR$_{c=1}$ vs TwitterRank & 0.4 \\
 TIR$_{c=0.5}$ vs TwitterRank & 0.65 \\ \hline
 \end{tabular}
\end{table}

\subsection{Friend Recommendation}
We further investigate the performance of TIR, TunkRank and TwitterRank with respect to friend recommendation (also known as the link prediction) by conducting the same experiment as in \citet{weng2010twitterrank}. As listed in Table \ref{table:selection}, links that need to be predicted are selected based on either friend attribute or similarities between the two users. Each of the eight link sets represents a specific test scenario and contains 30 links randomly selected from all following links regarding its selection criteria. $L_{fh}$, for example, we first rank all the links based on the number of followers of the friend user, then randomly select 30 links from the top $10\%$ links. Note that the Jenson-Shannon distance (one of the selection criteria in Table \ref{table:selection}) is calculated based on the feature vectors of the two users.

\begin{table*}[ht]
\centering
\renewcommand{\arraystretch}{1.2}
\caption{Selected test scenarios}
\label{table:selection}
\begin{tabular}{c|c|l} \hline
 Link Set & Rank User & Selection Criteria \\ \hline
 $L_{fh}$ & friend & the number of followers (high $10\%$) \\
 $L_{fl}$ & friend & the number of followers (low $10\%$)\\
  $L_{th}$ & friend & the number of tweets (high $10\%$) \\
 $L_{tl}$ & friend & the number of tweets (low $10\%$)\\
  $L_{dh}$ & both & Jenson-Shannon distance (high $10\%$) \\
$L_{dl}$ & both & Jenson-Shannon distance (low $10\%$) \\
  $L_{rr}$ & both & Followed each other \\
 $L_{ur}$ & both & Only the follower followed the friend\\ \hline
 \end{tabular}
\end{table*}

The experiment is carried out for each link set $L$ follows the steps below:
\begin{enumerate}
\item Take one link out of $L$: $l(u,v)$;
\item Randomly select 10 users who $u$ doesn't follow as the test candidate set $C$;
\item Remove link $l(u,v)$ from graph $G$ and denote the new graph by $G^{'}$;
\item Apply ranking model on $G^{'}$ and calculate:
\begin{equation*} \label{eq:perfromance}
Q(l)=\vert \lbrace v^{'} | v^{'}  \in C, Rank(v^{'}) > Rank(v) \rbrace\vert
\end{equation*}
\item Repeat 1 - 4 for all links in $L$ and calculate the average $Q(l)$.
\end{enumerate}

As the purpose is to recommend friends for $u$, the personal perspective influence ranking instead of the global ranking is applied here for both TwitterRank and TIR. Since the ground truth is ``$u$ follows $v$'', the higher the $Q(l)$ the better the performance. As illustrated in Figure \ref{fig:compare}, the TIR model demonstrates better performance and more stability than TunkRank and TwitterRank for most of the scenarios. More specifically, TIR achieves better performance than TwitterRank in 7 test scenarios and outperforms TunkRank in 6 scenarios, especially in $L_{fl}$ and $L_{tl}$ where TunkRank and TwitterRank both show the worst performance.

Observe that TIR is outperformed by TunkRank in $L_{fh}$. The reason is that TunkRank fully relies on the network structure which leaves it to the other extreme when recommending friends with few followers ($L_{fl}$). TwitterRank is better than TIR in $L_{rr}$ which suggests that users who followed each other share similar topic interests even though they may have few interactions.

The influence of the penalty factor $c$ varies between scenarios. As shown in Figure \ref{fig:friendRecommendation}, $c$ only has a significant influence in TIR when $c \in [0.95, 1.0]$. It increases the performance in recommending friends with few followers ($L_{fl}$) which indicates that it is likely that a user follows a friend who has few followers because he is a close friend. When recommending friends who either follow back ($L_{rr}$) or not ($L_{ur}$), a higher $c$ decreases the performance which means users follow each other are not necessarily likely to interact with each other. However, the number of tweets a friend has doesn't affect his interactions with followers as $c$ shows little impact in $L_{th}$ and $L_{tl}$.

\TwoFig{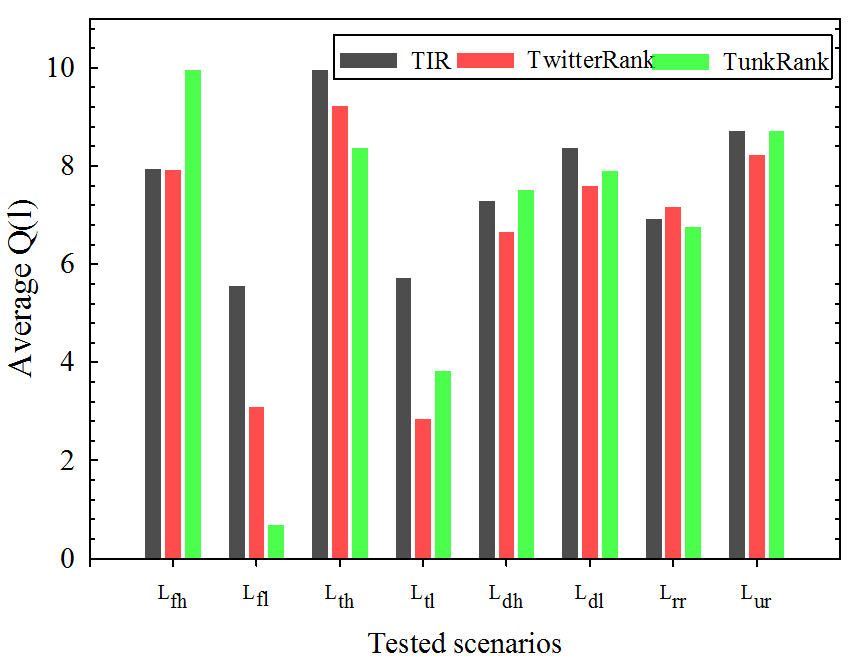}    {Performance in friend recommendation}    {fig:compare}
       {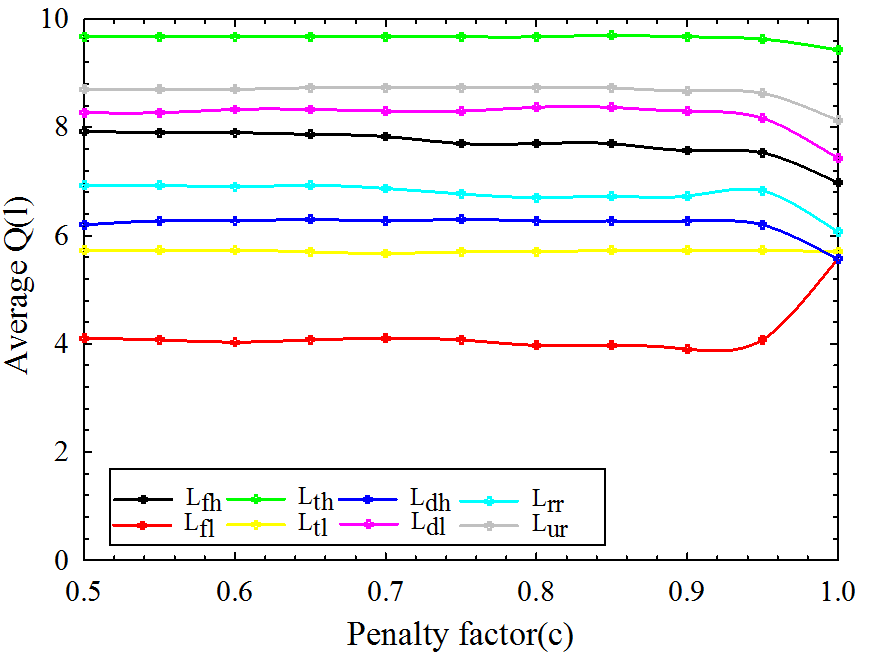}{The influence of penalty factor $c$ to TIR}    {fig:friendRecommendation}

\section{Conclusion and Future Work}
In this paper, we first investigated the temporal characteristics of user activity and found that user activities display certain patterns either for overall user activity or individual level activity. Moreover, we found that users tend to ignore tweets posted one hour ago and don't like to read many earlier tweets. Such characteristics suggest that user influence is time-sensitive and highly affected by user activity patterns. 

As response is an explicit indicator of user influence on others, we further formulated the response probability using a logistic regression model. Based on the formulated response probability, we proposed a PageRank based new influence ranking model (i.e., TIR) to estimate hourly user influence by taking hourly response probability as transition probability and introduced a penalty factor to bias the random walk towards close friends. Our method (TIR) was evaluated against two existing models: TunkRank and TwitterRank, in global influence ranking and friend recommendation separately. Experimental results demonstrated that TIR is more flexible than TunkRank and TwitterRank, and can better differentiate those influentials with more responsive followers. Results in friend recommendation substantiated the claim that TIR is more accurate and stable in a number of scenarios.

The next step of our research is to develop a web service to provide hourly based influentials for other applications. Meanwhile, TIR can also be used to develop useful extensions for Twitter such as finding the most influential friends for individual users or providing advice on when to post in order to attract more responses. Moreover, we will adapt our work to heterogeneous social networks so as to find influentials across multiple online social networks.


\bibliographystyle{agsm}    

\bibliography{AUSDM2016}  

\end{document}